\documentclass[twoside,draft,reqno]{birkart} 
 
\usepackage{amsfonts}

\input epsf

\newcommand{\etal}{\hbox{\it et al.}}
\newcommand{\Rset}{\mathbb{R}}

\begin{document} 
\setcounter{page}{1} 
 
\title[Secondary instabilities of hexagonal Faraday waves] 
{Secondary instabilities of hexagons: \\
a bifurcation analysis of experimentally \\
observed Faraday wave patterns}
\author[A.M.~Rucklidge]{A.M.~Rucklidge}
 
\address{%% 
Department of Applied Mathematics,\br 
University of Leeds, Leeds LS2 9JT UK}

\email{A.M.Rucklidge@leeds.ac.uk} 
 
\author[M.~Silber]{M.~Silber}

\address{%%
Department of Engineering Sciences and Applied Mathematics,\br
Northwestern University, Evanston IL 60208 USA}

\email{m-silber@northwestern.edu}

\author[J.~Fineberg]{J.~Fineberg}

\address{%%
The Racah Institute of Physics,\br
The Hebrew University of Jerusalem,\br
Givat Ram, Jerusalem 91904 Israel}

\email{jay@vms.huji.ac.il}

\begin{abstract} 
We examine three experimental observations of Faraday waves generated by
two-frequency forcing, in which a primary hexagonal pattern becomes
unstable to three different superlattice patterns. We use the
symmetry-based approach developed by Tse \etal~\cite{refT53} to
analyse the bifurcations involved in creating the three new
patterns. Each of the three examples reveals a different situation
that can arise in the theoretical analysis.
\end{abstract} 
 
\maketitle 
 
\section{Introduction} 
The classic Faraday wave experiment consists of a horizontal layer of fluid
that spontaneously develops a pattern of standing waves on its surface as it is
driven by vertical oscillation with amplitude exceeding a critical value. 
Recent experiments have revealed a wide variety of complex patterns, 
particularly in the large aspect ratio regime and with a forcing function 
containing two commensurate frequencies~\cite{refE10,refK84,refM118}. 
Transitions from the flat surface to a primary, spatially
periodic, pattern can be studied using equivariant bifurcation
theory~\cite{refG59}. These group theoretic techniques may also be
applied to secondary spatial period-multiplying transitions to
patterns with two distinct spatial scales (so called {\em
superlattice} patterns) as demonstrated by Tse \etal~\cite{refT53}.

We apply the method of Tse \etal~\cite{refT53} to the analysis of
three superlattice patterns observed when secondary subharmonic
instabilities destroy the basic hexagonal standing wave pattern in
two-frequency Faraday wave experiments.  We can make use not only of
the general symmetry-based approach from~\cite{refT53} but also of
many of the detailed results. The reason for this is that in
their paper, Tse \etal\ considered instabilities of hexagonal patterns
that broke the translation symmetry of the hexagons, but that remained
periodic in a larger hexagonal domain comprising twelve of the
original hexagons. The instabilities under consideration here satisfy
exactly the same conditions (though in fact they remain periodic in
smaller domains as well).
\looseness=-1

We begin by specifying the coordinate system and symmetries
we will use in section~\ref{sec:coordinates}, then describe the symmetries of
the three experimental patterns in section~\ref{sec:experimental}. In
section~\ref{sec:method}, we apply Tse \etal's method of analysis to these
three patterns, and present normal forms and stability calculations in
section~\ref{sec:normalforms}. We conclude in section~\ref{sec:discussion}.

 \begin{figure}
 \begin{center}
 \centerline{\epsfxsize3.0truein\epsffile{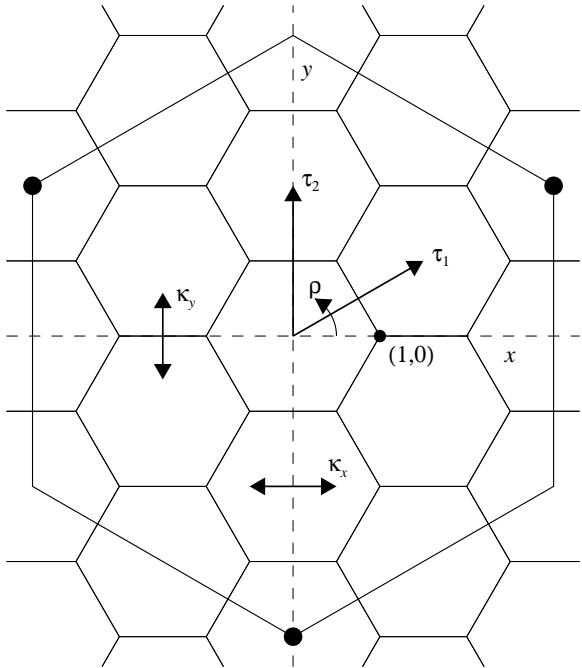}}
\caption{The coordinate system and certain elements of the symmetry
group~$\Gamma$. The origin of the coordinate system is at the centre
of the diagram, and the point $(1,0)$ is indicated. The small hexagons
represent the primary pattern, which is invariant under reflections
($\kappa_x$ and~$\kappa_y$), $60^\circ$~rotations ($\rho$) and
translations ($\tau_1$ and~$\tau_2$). The secondary patterns are all
periodic in the larger hexagonal box. The three corner points labelled
with solid circles are identified through the assumed periodicity.}
 \label{fig:coordinates}
 \end{center}
 \end{figure}
 
\section{Coordinates and symmetries}\label{sec:coordinates}
The primary pattern is made up of regular hexagons, which are invariant under
the group~$D_6$ (made up of $60^\circ$~rotations and reflections) combined with
translation from one hexagon to the next (see figure~\ref{fig:coordinates}).
Tse \etal~\cite{refT53} studied experimental patterns reported
in~\cite{refK67}, which had the feature that after the secondary instability,
the pattern remained periodic in the larger hexagonal box in
figure~\ref{fig:coordinates}. The 144-element spatial symmetry group of the
primary hexagonal pattern within this box is~$\Gamma$, generated by the
following reflection~$\kappa_x$, rotation~$\rho$ and translations $\tau_1$
and~$\tau_2$:
 \begin{alignat}{2}\label{eqn:symmetries}
 \kappa_x &: (x,y) \rightarrow (-x,y)                                 
          & \qquad
 \tau_1   &: (x,y) \rightarrow 
             (x,y) + \left(\frac{3}{2},\frac{\sqrt{3}}{2}\right)      \\
 \rho     &: (x,y) \rightarrow 
             \left(\frac{1}{2}x-\frac{\sqrt{3}}{2}y,
                   \frac{\sqrt{3}}{2}x+\frac{1}{2}y\right)              
          & \qquad
 \tau_2   &: (x,y) \rightarrow (x,y)+\left(0,\sqrt{3}\right)
 \end{alignat}
We also define $\kappa_y=\kappa_x\rho^3$, and note the following identities:
 \begin{gather}
 \kappa_x^2=\kappa_y^2=\rho^6=\tau_1^6=\tau_2^6=\tau_1^2\tau_2^2
           =\hbox{identity},                                          \\
 \rho\kappa_x = \kappa_x\rho^5,\qquad 
 \tau_1\kappa_x = \kappa_x\tau_1^5\tau_2,\qquad
 \tau_2\kappa_x = \kappa_x\tau_2,                                     \\
 \tau_1\rho = \rho\tau_1^3\tau_2,\qquad
 \tau_2\rho = \rho\tau_1,\qquad
 \tau_1\tau_2 = \tau_2\tau_1.
 \end{gather}
The time translation $\tau_T$ advances time by one period~$T$ of the forcing
function, which is the same as the temporal period of the hexagonal pattern.
This time translation is combined with the spatial symmetries above to give
spatio-temporal symmetries.

 \begin{figure}
 \begin{center}
 \centerline{%
 \hbox to0.30\hsize{\hfil(a)\hfil}\hfil
 \hbox to0.30\hsize{\hfil(b)\hfil}\hfil
 \hbox to0.30\hsize{\hfil(c)\hfil}}
 \centerline{% 
 \epsfxsize0.30\hsize\epsffile{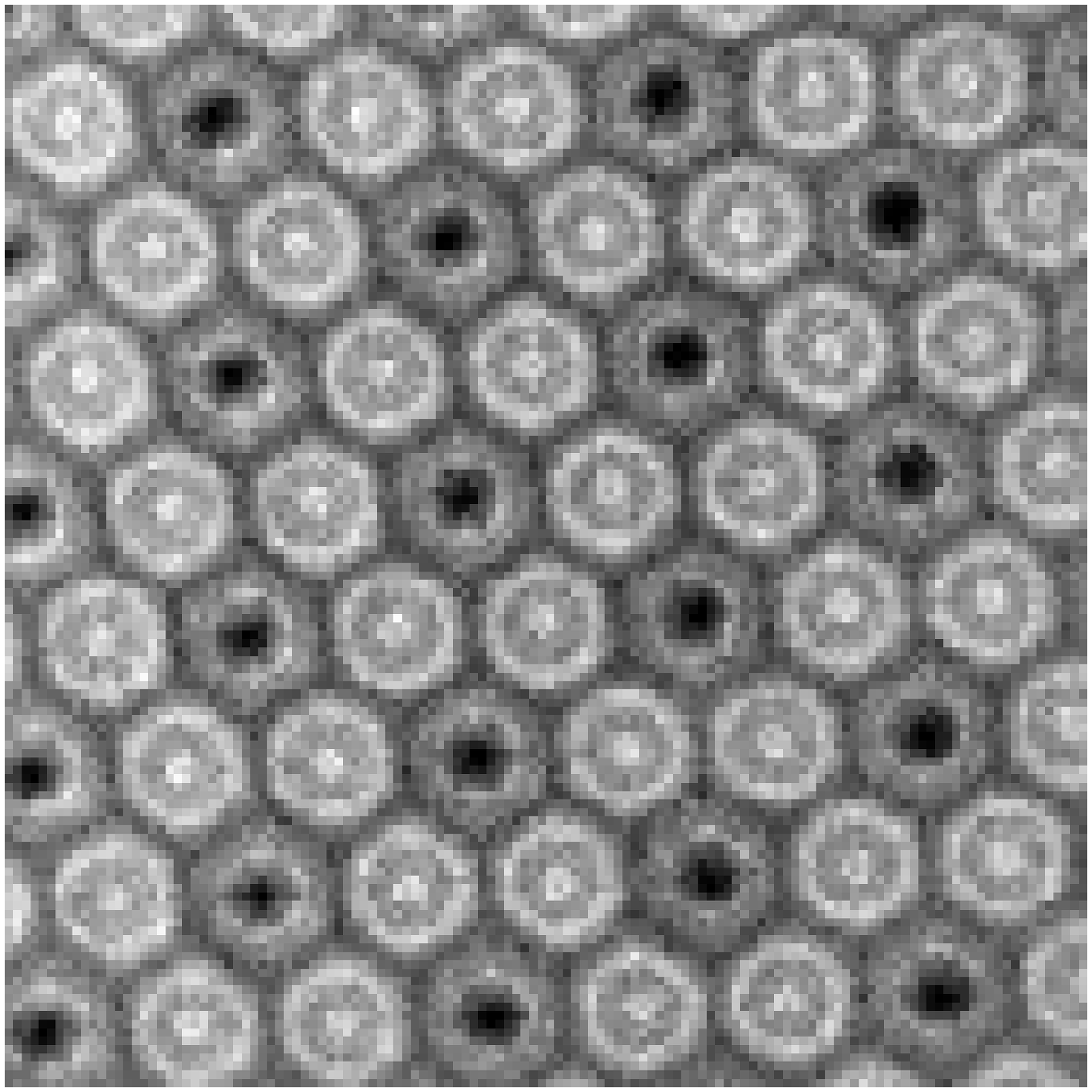}\hfil
 \epsfxsize0.30\hsize\epsffile{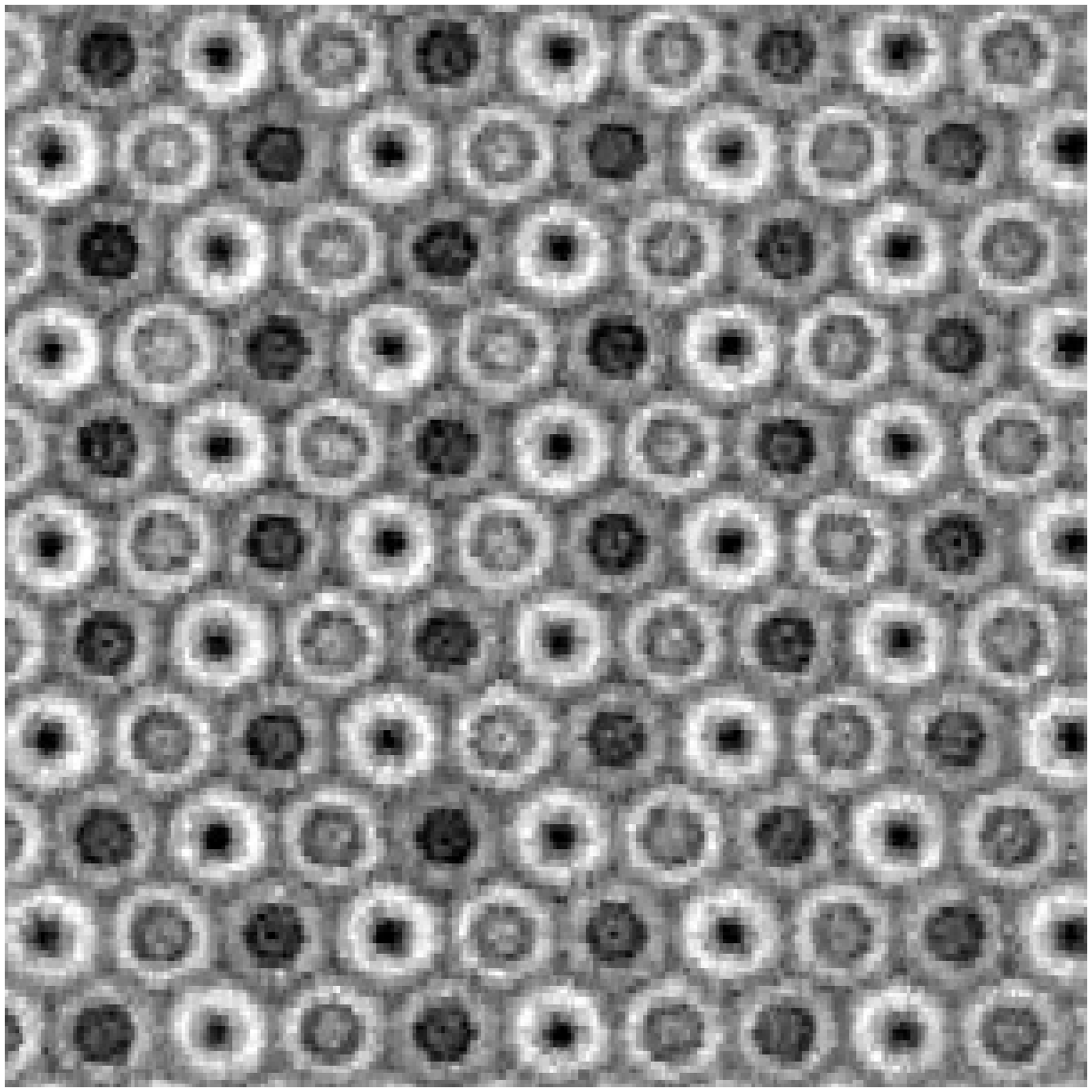}\hfil
 \epsfxsize0.30\hsize\epsffile{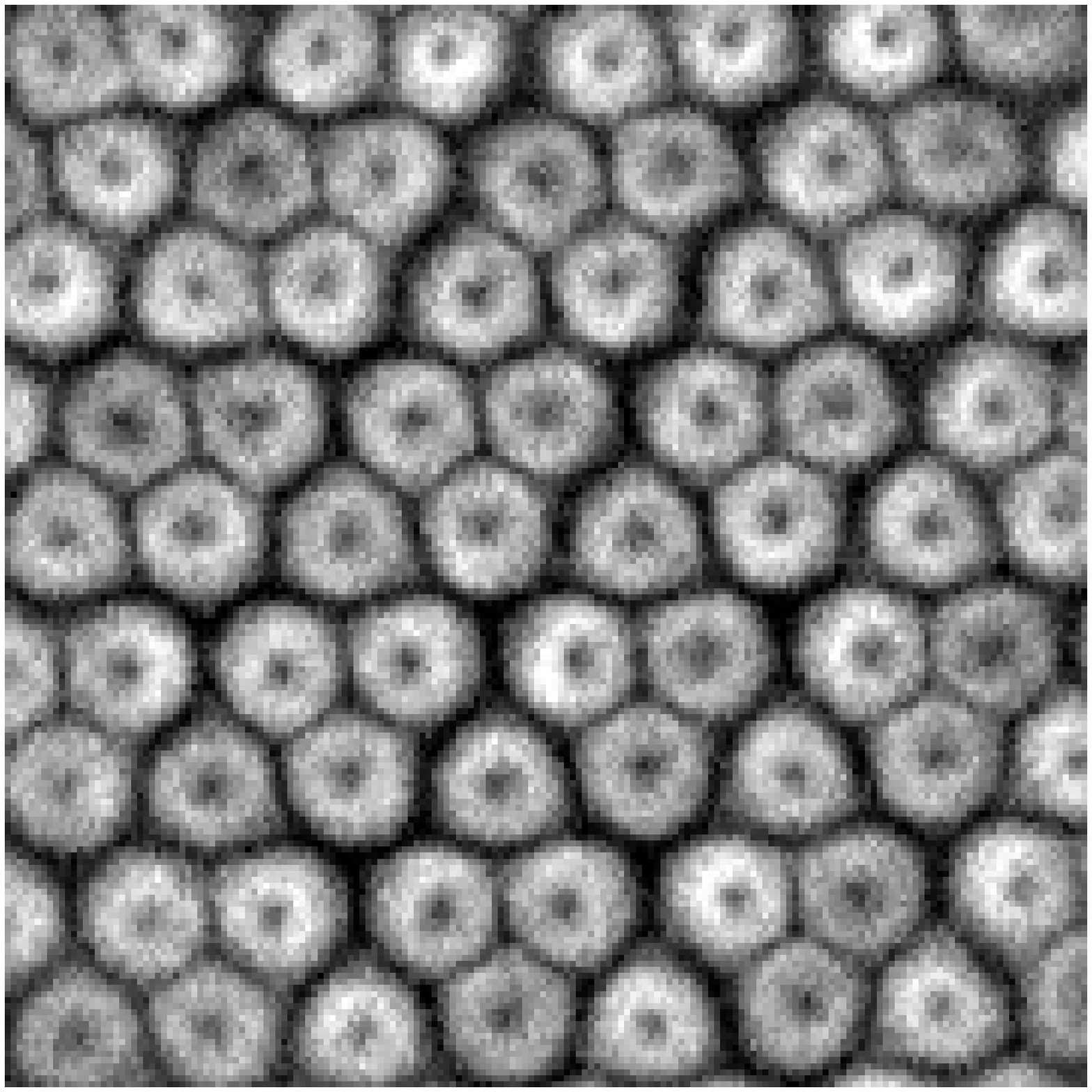}}
 \centerline{%
 \hbox to0.30\hsize{\hfil(d)\hfil}\hfil
 \hbox to0.30\hsize{\hfil(e)\hfil}\hfil
 \hbox to0.30\hsize{\hfil(f)\hfil}}
 \centerline{% 
 \epsfxsize0.30\hsize\epsffile{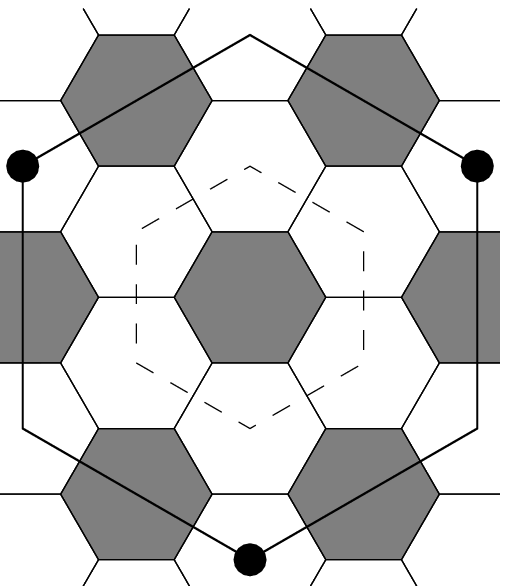}\hfil
 \epsfxsize0.30\hsize\epsffile{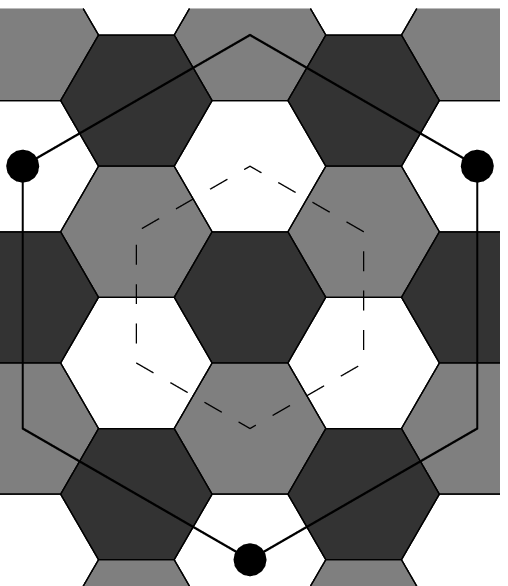}\hfil
 \epsfxsize0.30\hsize\epsffile{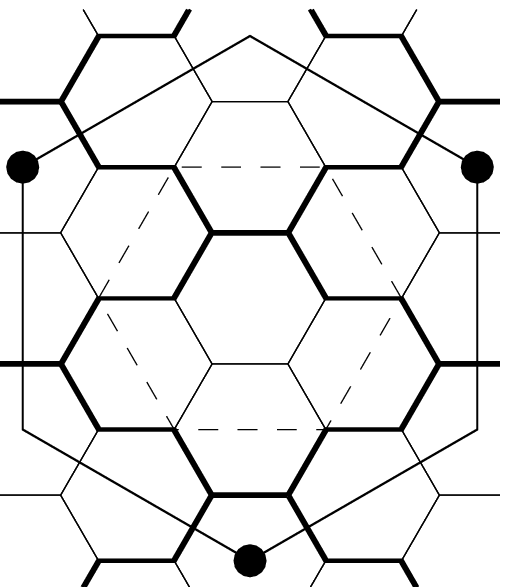}}
 \centerline{%
 \hbox to0.30\hsize{\hfil  \hfil}\hfil
 \hbox to0.30\hsize{\hfil(g)\hfil}\hfil
 \hbox to0.30\hsize{\hfil(h)\hfil}}
 \centerline{% 
 \hbox to 0.30\hsize{\hfil}\hfil
 \epsfxsize0.30\hsize\epsffile{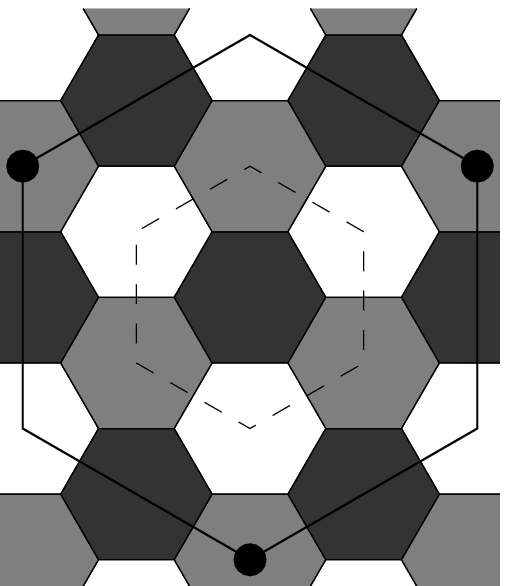}\hfil
 \epsfxsize0.30\hsize\epsffile{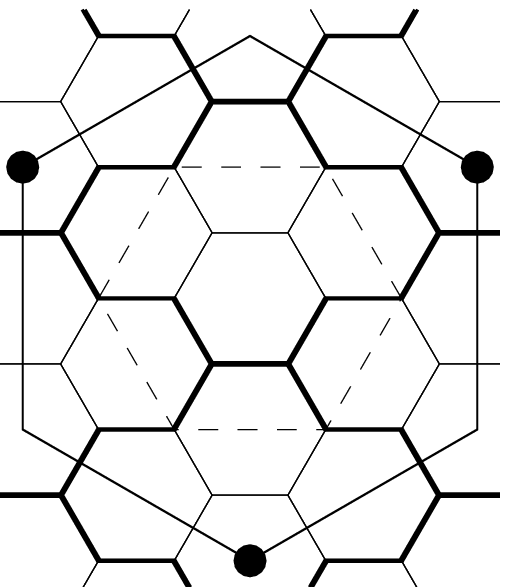}}
 \caption{Experimental and idealised secondary patterns. (a-c)~Experimental 
 patterns, visualised from above. (d-f)~Idealised versions of~(a-c). 
 (g-h)~patterns (e-f) but seen one forcing period~$T$ later. The idealisations 
 are all rotated by about~$30^\circ$ compared with the experimental pictures.}
 \label{fig:idealised}
 \end{center}
 \end{figure}

\section{Experimental patterns}\label{sec:experimental}
The three experimentally observed patterns are shown in
figure~\ref{fig:idealised}(a-c), visualised using the techniques described
in~\cite{refA57}. Patterns (a) and (b) are both obtained using Dow-Corning
silicone oil with viscosity $47\,\hbox{cSt}$ and layer depth $0.35\,\hbox{cm}$,
while pattern (c) was found using a $23\,\hbox{cSt}$ oil layer of depth
$0.155\,\hbox{cm}$.  All three patterns are obtained with forcing function
containing two frequencies in the ratio $2:3$; pattern~(a) is found with
frequencies $50$ and $75\,\hbox{Hz}$, pattern~(b) with frequencies $70$ and
$105\,\hbox{Hz}$, and pattern~(c) with $40$ and $60\,\hbox{Hz}$ driving
frequencies. Pattern (c) was reported previously in~\cite{refA57}. Typically,
the secondary bifurcations occur at forcing amplitudes between 10 and 50\%
larger than the critical acceleration for the primary hexagonal state. Further
experimental details can be found in~\cite{refA57,refL61}.

For the purposes of the analysis, we consider the idealised versions of these
experimental patterns, shown in figure~\ref{fig:idealised}(d-f). The first
pattern in figure~\ref{fig:idealised}(a,d) retains the $D_6$~symmetry of the
original hexagons but breaks certain translation symmetries.  It is periodic in
the medium-sized dashed hexagon in figure~\ref{fig:idealised}(d), which implies
that the pattern is invariant under the translations~$\tau_1^3$
and~$\tau_1\tau_2$. It has no spatio-temporal symmetries. The second pattern is
similar, although it possesses only triangular ($D_3$) symmetry
instantaneously. Moreover, it has the spatio-temporal symmetry given by a
$60^\circ$ rotation combined with advance in time by one period~$T$ of the
forcing, as in figure~\ref{fig:idealised}(e,g). In fact, this spatio-temporal
symmetry was first suggested by the analysis below, and found to be consistent
with the experimental observations. The third pattern in
figure~\ref{fig:idealised}(c,f) is quite different: the dark lozenges in
figure~\ref{fig:idealised}(f) represent the enlarged gaps between the hexagons
in figure~\ref{fig:idealised}(c). The pattern is periodic in the medium-sized
dashed hexagon in figure~\ref{fig:idealised}(f), so is invariant under
translations $\tau_1^2$ and~$\tau_2^2=\tau_1^4$. It is also invariant under the
group of symmetries of a rectangle $D_2$, and possesses the spatio-temporal
symmetry of the translation~$\tau_2$ combined with advance in time by one
period~$T$ of the forcing, as in figure~\ref{fig:idealised}(f,h).

Using the information above, we write down the instantaneous (spatial) symmetry
groups of the three patterns from figure~\ref{fig:idealised}(a-c) in terms of
their generators:
 \begin{equation}
 \Sigma_a=\langle\kappa_x,\rho,\tau_1^3,\tau_1\tau_2\rangle,\quad
 \Sigma_b=\langle\kappa_x,\rho^2,\tau_1^3,\tau_1\tau_2\rangle,\quad
 \Sigma_c=\langle\kappa_x,\kappa_y\tau_2,\tau_1^2\rangle.
 \end{equation}
These groups are of order 48, 24 and 12 respectively. For the full
spatio-temporal symmetry groups, we would also include $\rho\tau_T$ in the
generators of~$\Sigma_b$, and $\tau_2\tau_T$ in the generators of~$\Sigma_c$,
but initially we will work with the spatial symmetry groups. The reason for
this is that the instantaneous (spatial) symmetries can be determined reliably
from a single experimental image, while extracting spatio-temporal symmetries
from the experimental data is more involved.

% needed to adjust the text size (small) and intercolumn spacing (8pt)
% to fit into the specified style.

\begin{table}
\begin{center}
\small
\begin{tabular}{r|r@{\hspace{8pt}}r@{\hspace{8pt}}r@{\hspace{8pt}}
                  r@{\hspace{8pt}}r@{\hspace{8pt}}r@{\hspace{8pt}}
                  r@{\hspace{8pt}}r@{\hspace{8pt}}r@{\hspace{8pt}}
                  r@{\hspace{8pt}}r@{\hspace{8pt}}r@{\hspace{8pt}}
                  r@{\hspace{8pt}}r@{\hspace{8pt}}r} 
\hline 
             & a & b & c & d & e & f & g & h & i & j & k & l & m & n & o \\
                    &
id                  &
$\kappa_x$          &
$\kappa_y$          &
$\tau_1$            &
$\tau_1^2$          &
$\tau_1^3$          &
$\kappa_x\tau_1$    &
$\kappa_x\tau_2$    &
$\kappa_x\tau_1^3$  &
$\kappa_y\tau_1^3$  &
$\rho$              &
$\rho^2$            &
$\rho^3$            &
$\rho^2\tau_1$      &
$\rho^3\tau_1^3$    \\
   & 1 & 6    & 18   & 6    & 2    & 3    & 12   & 12   & 6    & 18   & 24   & 8    & 3    & 16   & 9 \\
\hline
%    a     b      c      d      e      f      g      h      i      j      k      l      m      n      o
 1 & 1 &  $1$ &  $1$ &  $1$ &  $1$ &  $1$ &  $1$ &  $1$ &  $1$ &  $1$ &  $1$ &  $1$ &  $1$ &  $1$ &  $1$ \\ 
 2 & 1 & $-1$ & $-1$ &  $1$ &  $1$ &  $1$ & $-1$ & $-1$ & $-1$ & $-1$ &  $1$ &  $1$ &  $1$ &  $1$ &  $1$ \\
 3 & 1 &  $1$ & $-1$ &  $1$ &  $1$ &  $1$ &  $1$ &  $1$ &  $1$ & $-1$ & $-1$ &  $1$ & $-1$ &  $1$ & $-1$ \\
 4 & 1 & $-1$ &  $1$ &  $1$ &  $1$ &  $1$ & $-1$ & $-1$ & $-1$ &  $1$ & $-1$ &  $1$ & $-1$ &  $1$ & $-1$ \\
 5 & 2 &  $0$ &  $0$ &  $2$ &  $2$ &  $2$ &  $0$ &  $0$ &  $0$ &  $0$ &  $1$ & $-1$ & $-2$ & $-1$ & $-2$ \\
 6 & 2 &  $0$ &  $0$ &  $2$ &  $2$ &  $2$ &  $0$ &  $0$ &  $0$ &  $0$ & $-1$ & $-1$ &  $2$ & $-1$ &  $2$ \\
 7 & 2 &  $2$ &  $0$ & $-1$ & $-1$ &  $2$ & $-1$ & $-1$ &  $2$ &  $0$ &  $0$ &  $2$ &  $0$ & $-1$ &  $0$ \\
 8 & 2 & $-2$ &  $0$ & $-1$ & $-1$ &  $2$ &  $1$ &  $1$ & $-2$ &  $0$ &  $0$ &  $2$ &  $0$ & $-1$ &  $0$ \\
 9 & 3 &  $1$ &  $1$ & $-1$ &  $3$ & $-1$ & $-1$ &  $1$ & $-1$ & $-1$ &  $0$ &  $0$ &  $3$ &  $0$ & $-1$ \\
10 & 3 & $-1$ &  $1$ & $-1$ &  $3$ & $-1$ &  $1$ & $-1$ &  $1$ & $-1$ &  $0$ &  $0$ & $-3$ &  $0$ &  $1$ \\
11 & 3 & $-1$ & $-1$ & $-1$ &  $3$ & $-1$ &  $1$ & $-1$ &  $1$ &  $1$ &  $0$ &  $0$ &  $3$ &  $0$ & $-1$ \\
12 & 3 &  $1$ & $-1$ & $-1$ &  $3$ & $-1$ & $-1$ &  $1$ & $-1$ &  $1$ &  $0$ &  $0$ & $-3$ &  $0$ &  $1$ \\
13 & 4 &  $0$ &  $0$ & $-2$ & $-2$ &  $4$ &  $0$ &  $0$ &  $0$ &  $0$ &  $0$ & $-2$ &  $0$ &  $1$ &  $0$ \\
14 & 6 & $-2$ &  $0$ &  $1$ & $-3$ & $-2$ & $-1$ &  $1$ &  $2$ &  $0$ &  $0$ &  $0$ &  $0$ &  $0$ &  $0$ \\
15 & 6 &  $2$ &  $0$ &  $1$ & $-3$ & $-2$ &  $1$ & $-1$ & $-2$ &  $0$ &  $0$ &  $0$ &  $0$ &  $0$ &  $0$ \\
\hline
\end{tabular}
\vspace{2mm}
 \caption{Character table of the group~$\Gamma$, taken from Tse 
 \etal, with corrections. A representative element is shown on the second line
 for each conjugacy class (see also figure~\ref{fig:classes}),
 and the number of elements in the class is on the 
 third row. The next fifteen rows give the characters associated with 
 each conjugacy class for each of the fifteen representations.}
 \label{tab:chartable}
 \end{center}
 \end{table}

Each of the three instabilities that generates the three different patterns
will be associated with a set of marginally stable eigenfunctions; the new
pattern, at least near onset, can be thought of as (approximately) a linear
combination of these marginal eigenfunctions and the original hexagonal
pattern. Which linear superpositions are consistent with the nonlinearity
inherent in the pattern formation process is determined by our bifurcation
analysis. The symmetries in~$\Gamma$ all leave the primary hexagonal pattern
unchanged, so they must send marginal eigenfunctions onto linear combinations
of marginal eigenfunctions, which induces an action on the amplitudes of these
functions. In other words, if there are $n$~marginal eigenfunctions $f_1$,
\dots,~$f_n$, with $n$~amplitudes $\mathbf{a}=(a_1,\dots,a_n)\in\Rset^n$, each
element $\gamma\in\Gamma$ sends $\mathbf{a}$ to $R_\gamma\mathbf{a}$, where the
set of $n\times n$ orthogonal matrices $R_\gamma$ forms a
representation~$R_\Gamma$ of the group~$\Gamma$. For
subharmonic instabilities of the type of interest here, this will generically
be an irreducible representation (irrep)~\cite{refG59}. Tse \etal~\cite{refT53}
have computed all the irreps of the group~$\Gamma$; the character table of
these representations is reproduced in table~\ref{tab:chartable}. Recall that
the character of a group element~$\gamma$ in a representation is the trace of
the matrix~$R_\gamma$, and that conjugate elements (which form a conjugacy
class) have the same characters.

Once the representation associated with each of the three transitions is
identified, we can write down the normal form, work out what other patterns are
created in the same bifurcation, and compute stability of the patterns in terms
of the normal form coefficients.

 \begin{figure}
 \begin{center}
 \centerline{%
 \hbox to0.191\hsize{\hfil(a) identity (1)\hfil}\hfil
 \hbox to0.191\hsize{\hfil(b) $\kappa_x$ (6)\hfil}\hfil
 \hbox to0.191\hsize{\hfil(c) $\kappa_y$ (18)\hfil}}
 \centerline{% 
 \epsfxsize0.191\hsize\epsffile{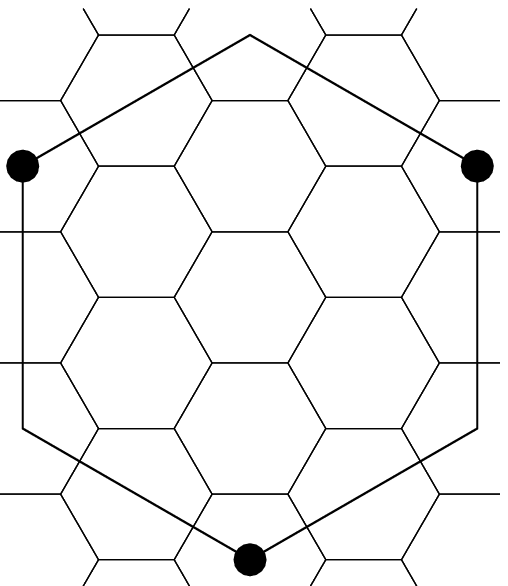}\hfil
 \epsfxsize0.191\hsize\epsffile{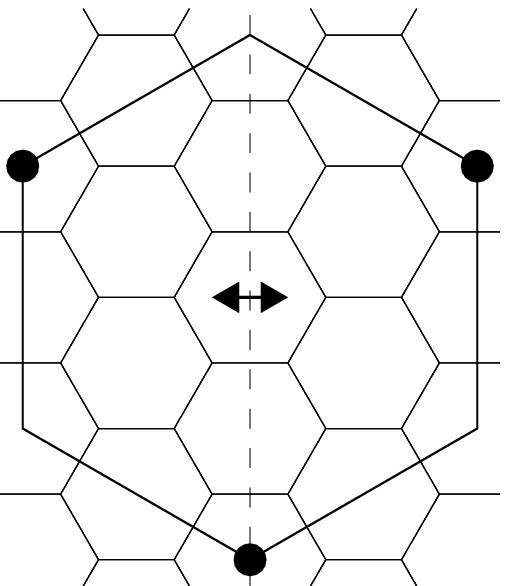}\hfil
 \epsfxsize0.191\hsize\epsffile{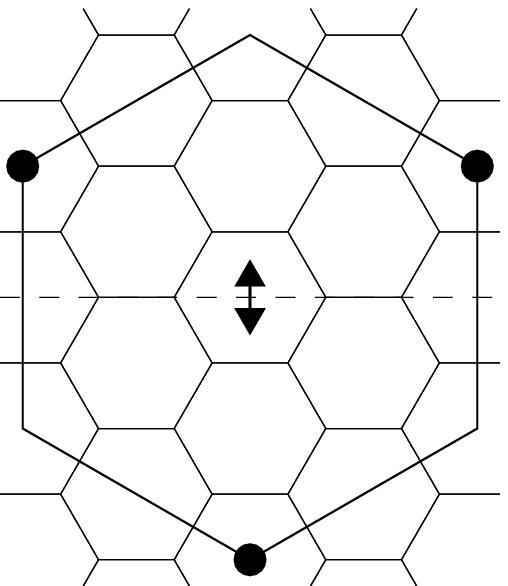}}
 \centerline{%
 \hbox to0.191\hsize{\hfil(d) $\tau_1$ (6)\hfil}\hfil
 \hbox to0.191\hsize{\hfil(e) $\tau_1^2$ (2)\hfil}\hfil
 \hbox to0.191\hsize{\hfil(f) $\tau_1^3$ (3)\hfil}}
 \centerline{% 
 \epsfxsize0.191\hsize\epsffile{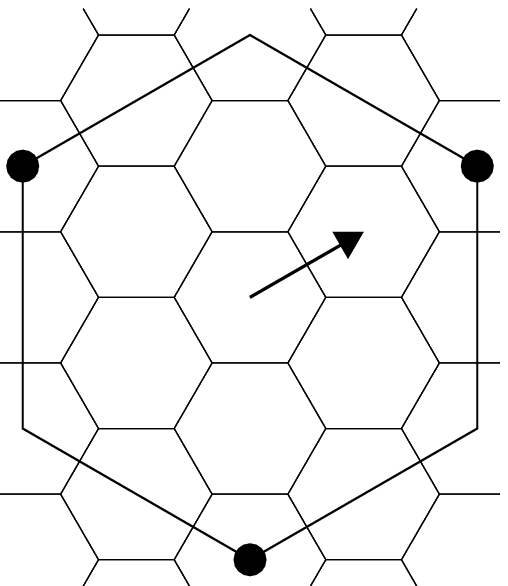}\hfil
 \epsfxsize0.191\hsize\epsffile{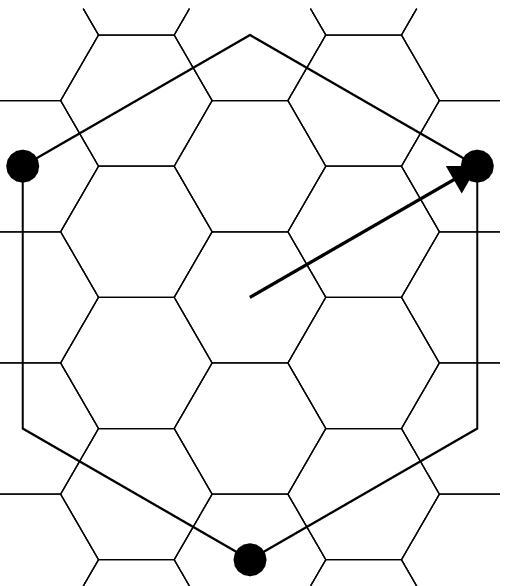}\hfil
 \epsfxsize0.191\hsize\epsffile{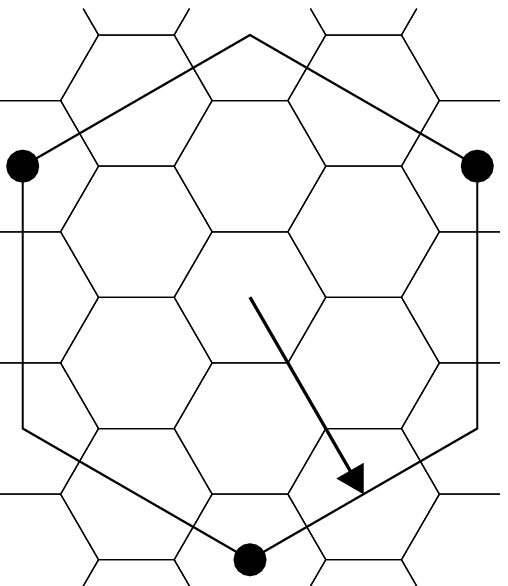}}
 \centerline{%
 \hbox to0.191\hsize{\hfil(g) $\kappa_x\tau_1$ (12)\hfil}\hfil
 \hbox to0.191\hsize{\hfil(h) $\kappa_x\tau_2$ (12)\hfil}\hfil
 \hbox to0.191\hsize{\hfil(i) $\kappa_x\tau_1^3$ (6)\hfil}}
 \centerline{% 
 \epsfxsize0.191\hsize\epsffile{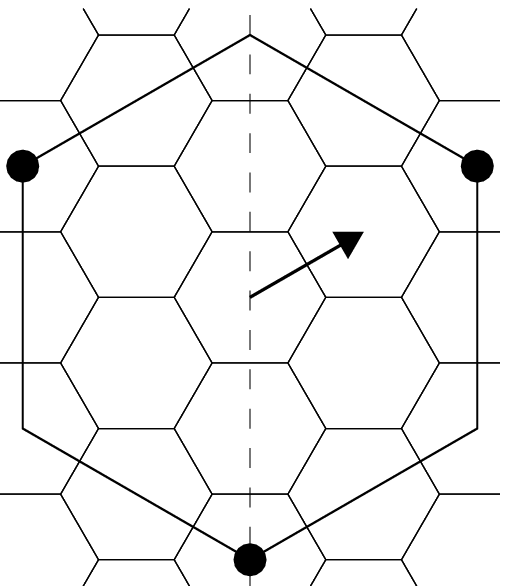}\hfil
 \epsfxsize0.191\hsize\epsffile{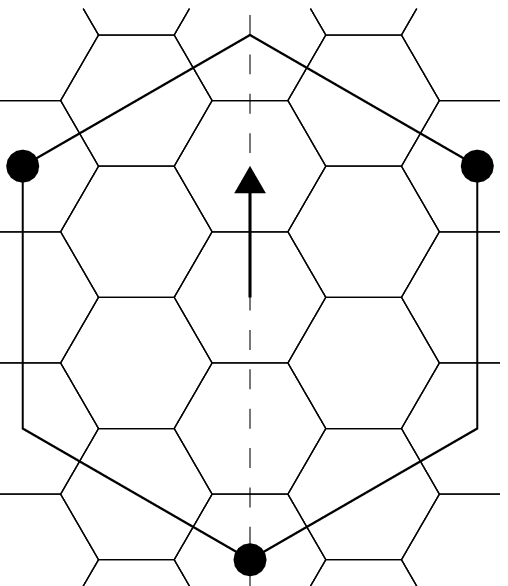}\hfil
 \epsfxsize0.191\hsize\epsffile{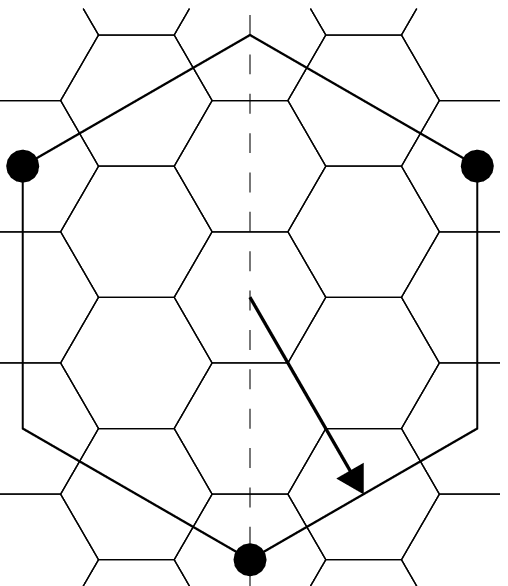}}
 \centerline{%
 \hbox to0.191\hsize{\hfil(j) $\kappa_y\tau_1^3$ (18)\hfil}\hfil
 \hbox to0.191\hsize{\hfil(k) $\rho$ (24)\hfil}\hfil
 \hbox to0.191\hsize{\hfil(l) $\rho^2$ (8)\hfil}}
 \centerline{% 
 \epsfxsize0.191\hsize\epsffile{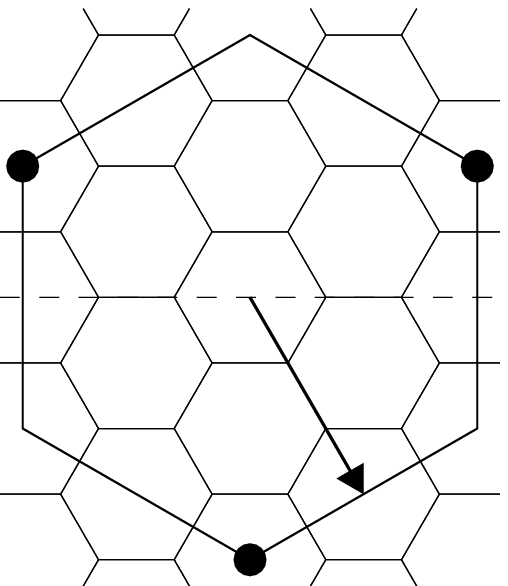}\hfil
 \epsfxsize0.191\hsize\epsffile{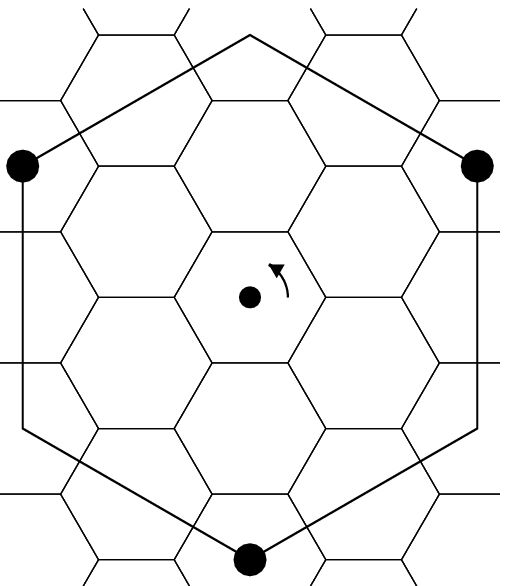}\hfil
 \epsfxsize0.191\hsize\epsffile{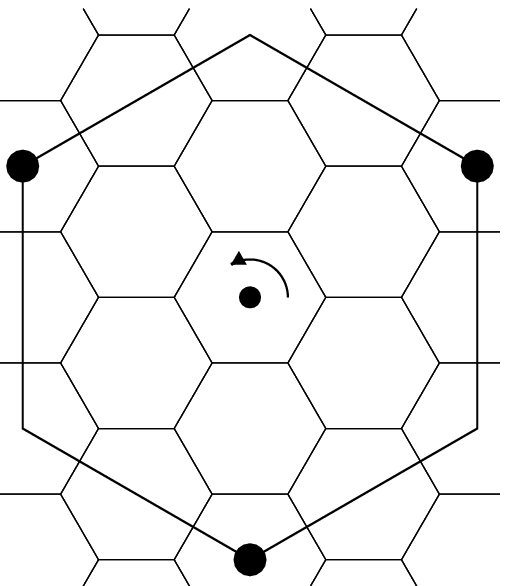}}
 \centerline{%
 \hbox to0.191\hsize{\hfil(m) $\rho^3$ (3)\hfil}\hfil
 \hbox to0.191\hsize{\hfil(n) $\rho^2\tau_1$ (16)\hfil}\hfil
 \hbox to0.191\hsize{\hfil(o) $\rho^3\tau_1^3$ (9)\hfil}}
 \centerline{% 
 \epsfxsize0.191\hsize\epsffile{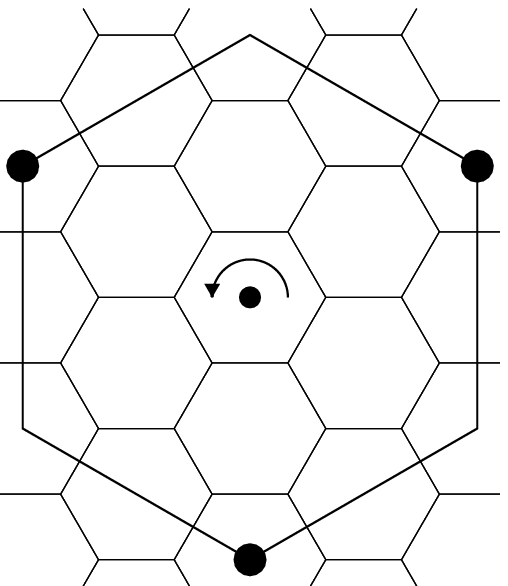}\hfil
 \epsfxsize0.191\hsize\epsffile{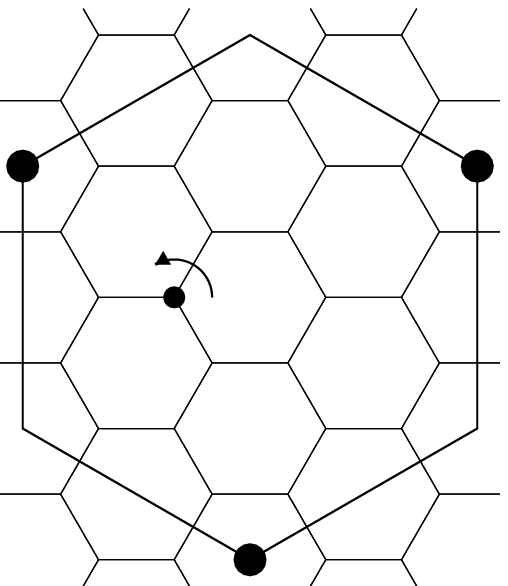}\hfil
 \epsfxsize0.191\hsize\epsffile{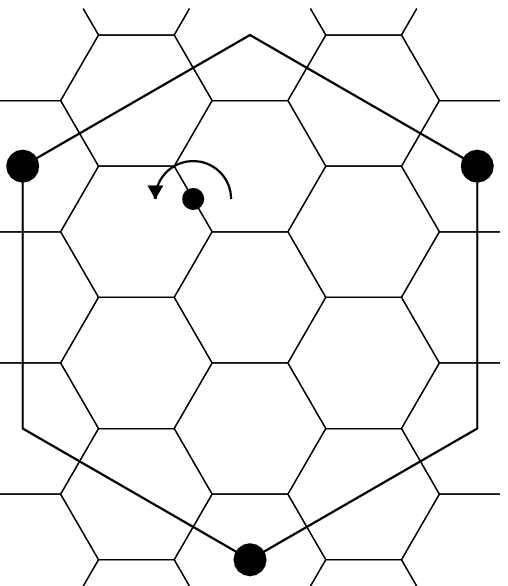}}
 \caption{The 15 conjugacy classes of~$\Gamma$. One 
 element from and the number of elements in each class are indicated. The 
letters (a)--(o) correspond to the columns of table~\ref{tab:chartable}.}
 \label{fig:classes}
 \end{center}
 \end{figure}

\section{Method}\label{sec:method}
The first task is to identify which representation is relevant for each
bifurcation. Tse \etal~\cite{refT53} outlined a two-stage method to accomplish
this. First, any symmetry element that is represented by the identity matrix in
a particular representation must appear in the symmetry group of every branch
of solutions created in a  bifurcation with that representation. This can be
used to eliminate from consideration any representation that has an element
with character equal to the character of the identity that does not appear in
the symmetry group of the observed pattern. Second, we make use of the trace
formula from~\cite{refG59}, which gives the dimension of the subspace of
$\Rset^n$ that is fixed by a particular isotropy subgroup~$\Sigma$ of~$\Gamma$
with representation given by the matrices~$R_\Gamma$:
 \begin{equation}\label{eq:traceformula}
 \hbox{dim}\,\hbox{fix}(\Sigma) = 
   \frac{1}{|\Sigma|}\sum_{\sigma\in\Sigma}\hbox{Tr}\,R_\sigma,
 \end{equation}
where $|\Sigma|$ is the number of elements in~$\Sigma$. Specifically, we use
the trace formula to eliminate those representations for which the spatial
symmetry group of the pattern fixes a zero-dimensional subspace (implying that
the subgroup is not an isotropy subgroup); only the remaining representations
need be examined in more detail.
\looseness=-1

We proceed by first counting the number of elements in each conjugacy class for
each of the symmetry groups $\Sigma_a$, $\Sigma_b$ and~$\Sigma_c$.
Figure~\ref{fig:classes} shows representative elements from each class and is
helpful for this categorization. The result of this is:
 $\Sigma_a$ contains:
 \begin{equation}
 a:1,\, b:6,\, c:6,\, f:3,\, i:6,\, j:6,\, k:8,\, l:8,\, m:1,\, o:3
 \end{equation}
 (that is, one element from class~$a$, six from class~$b$ etc.); $\Sigma_b$
 contains:
 \begin{equation}
 a:1,\, b:6,\, f:3,\, i:6,\, l:8;
 \end{equation}
 and $\Sigma_c$ contains:
 \begin{equation}
 a:1,\, b:1,\, c:3,\, e:2,\, h:2,\, o:3.
 \end{equation}
The element $\tau_1^2$ does not appear in the symmetry groups of patterns (a)
and (b), which eliminates representations 1--6 and 9--12 (since $\tau_1^2$ is
represented by the identity matrix in all these: see
table~\ref{tab:chartable}). Similarly, $\tau_1\tau_2$ in class~$f$ and $\rho^3$
do not appear in $\Sigma_c$, which eliminates representations 1--9, 11 and 13
from consideration for that bifurcation problem.

Next, by applying~(\ref{eq:traceformula}), we find that pattern~(a) has a
non-zero dimensional fixed point subspace only in representation~7, as does
pattern~(b). The spatial symmetry group of pattern~(a) fixes a one-dimensional
subspace, and that of pattern~(b) fixes a two-dimensional subspace. Pattern~(c)
has a one-dimensional fixed point subspace in representations 10 and~12, and
zero in other representations.

We are therefore faced with three different situations: the spatial symmetry
group~$\Sigma_a$ fixes a one-dimensional subspace in representation~7, so we
expect by the Equivariant Branching Lemma (see~\cite{refG59}) that such a
pattern will generically be found in a bifurcation problem with that
representation.

Pattern~(b), on the other hand, has a spatial symmetry group that fixes a
two-dimensional subspace. However, we must take into account that the pattern
arises in a subharmonic (period-doubling) instability, and extend the
groups~$\Gamma$ and~$\Sigma_b$ to the spatio-temporal symmetry groups that
arise by including time translations. We may then show that the spatio-temporal
symmetry group of pattern~(b) fixes a one-dimensional subspace, and so also
arises generically in a subharmonic bifurcation with representation~7. This is
the same representation as with pattern~(a), obtained for similar experimental
parameter values. Extending to include the subharmonic nature of the
instability does not affect the branching of pattern~(a).

The third situation arises with pattern~(c), which on symmetry arguments alone
could be associated with either representation~10 or representation~12.
Including information about the spatio-temporal symmetry of the pattern does
not distinguish between these two representations. However, information on the
Fourier transform of the pattern does allow a choice to be made between the two
possibilities; in order to show this, we first need to work out which
combinations of Fourier modes are associated with each pattern.

It is useful to have sample Fourier modes for the basic hexagonal 
pattern:
 \begin{equation}\label{eq:basichexgons}
 f_0(x,y)=    \cos2\pi\left(\frac{2x}{3}\right) +
              \cos2\pi\left(-\frac{x}{3}+\frac{y}{\sqrt{3}}\right) +
              \cos2\pi\left(-\frac{x}{3}-\frac{y}{\sqrt{3}}\right),
 \end{equation}
with wavevector of length $\frac{4\pi}{3}$, as well as sample Fourier modes
for representations~7, 10 and~12. The method described by Tse
\etal~\cite{refT53} yields Fourier functions that would be included in the
eigenfunctions associated with representation~7; representative functions with
the shortest wavevectors include:
 \begin{align}\label{eq:repn7fns1}
 f_1(x,y)&=    \cos2\pi\left(\frac{x}{3}+\frac{y}{3\sqrt{3}}\right) +
               \cos2\pi\left(\frac{x}{3}-\frac{y}{3\sqrt{3}}\right) +
               \cos2\pi\left(\frac{2y}{3\sqrt{3}}\right)   \\
 f_2(x,y)&=    \sin2\pi\left(\frac{x}{3}+\frac{y}{3\sqrt{3}}\right) +
               \sin2\pi\left(-\frac{x}{3}+\frac{y}{3\sqrt{3}}\right) +
               \sin2\pi\left(-\frac{2y}{3\sqrt{3}}\right),
 \label{eq:repn7fns2}
 \end{align}
which is made up of wavevectors of length equal to $\frac{1}{\sqrt{3}}$ of that
of the basic hexagonal pattern. Eigenfunctions for representation~10 are made
up of Fourier functions that include:
 \begin{equation}\label{eq:repn10fns}
 f_1=\sin2\pi\left(\frac{x}{6}+\frac{y}{2\sqrt{3}}\right)\quad
 f_2=\sin2\pi\left(\frac{-x}{6}+\frac{y}{2\sqrt{3}}\right)\quad
 f_3=\sin2\pi\left(\frac{-x}{3}\right),
 \end{equation}
with wavevector of length $\frac{1}{2}$ the fundamental; and representation~12
has:
 \begin{equation}\label{eq:repn12fns}
 f_1=\sin2\pi\left(\frac{x}{2}+\frac{-y}{2\sqrt{3}}\right)\quad
 f_2=\sin2\pi\left(\frac{x}{2}+\frac{y}{2\sqrt{3}}\right)\quad
 f_3=\sin2\pi\left(\frac{y}{\sqrt{3}}\right),
 \end{equation}
with wavevector of length $\frac{\sqrt{3}}{2}$ the fundamental. In each case,
we have chosen the Fourier modes with the shortest wavevectors, as these are
easiest to identify in an experimental Fourier transform. 

The images of the Fourier transform of pattern~(c) in~\cite{refA57} show that
the mode created in the instability contains wavevectors that are a factor of~2
shorter than the shortest in the basic hexagonal pattern, which is consistent
with representation~10 but not~12. In this way, information about the power
spectrum of the pattern is necessary to supplement the arguments based
entirely on symmetries and to distinguish between the two choices.

\section{Normal forms}\label{sec:normalforms}
Using the functions specified above as a basis for representations~7 and~10,
the matrices that generate the two relevant representations are, for
representation~7:
 \begin{equation}
 R_{\kappa_x}= I_2,    \quad
 R_\rho=       \begin{bmatrix}1&0\\0&-1\end{bmatrix},   \quad
 R_{\tau_1}=   \begin{bmatrix}-\frac{1}{2}        &  \frac{\sqrt{3}}{2}\\
                              -\frac{\sqrt{3}}{2} & -\frac{1}{2} \end{bmatrix},   \quad
 R_{\tau_2}=   R_{\tau_1}^2,  \quad
 R_{\tau_T}=   -I_2, 
 \end{equation}
where $I_n$ is the $n\times n$ identity matrix; 
and for representation~10:
 \def\AMRpone{\phantom{-}1}
 \def\AMRmone{         - 1}
 \def\AMRzero{\phantom{-}0}
 \begin{align}
 R_{\kappa_x}&=  \begin{bmatrix}\AMRzero&\AMRpone&\AMRzero\\\AMRpone&\AMRzero&\AMRzero\\\AMRzero&\AMRzero&\AMRmone\end{bmatrix},  \qquad
 R_\rho=         \begin{bmatrix}\AMRzero&\AMRzero&\AMRmone\\\AMRpone&\AMRzero&\AMRzero\\\AMRzero&\AMRpone&\AMRzero\end{bmatrix},  \\
 R_{\tau_1}&=    \begin{bmatrix}\AMRmone&\AMRzero&\AMRzero\\\AMRzero&\AMRpone&\AMRzero\\\AMRzero&\AMRzero&\AMRmone\end{bmatrix},  \qquad
 R_{\tau_2}=     \begin{bmatrix}\AMRmone&\AMRzero&\AMRzero\\\AMRzero&\AMRmone&\AMRzero\\\AMRzero&\AMRzero&\AMRpone\end{bmatrix},  \qquad
 R_{\tau_T}=     -I_3.
 \end{align}
The perturbation amplitude at time $j+1$ times the forcing period, given the
perturbation at time~$j$, is given by
$\mathbf{a}_{j+1}=\mathbf{f}(\mathbf{a}_j)$, where the equivariance condition
amounts to $R_\gamma\mathbf{f}(\mathbf{a})=\mathbf{f}(R_\gamma\mathbf{a})$ for
all $\gamma\in\Gamma$. Using this, we can determine the relevant normal form
associated with these two representations:
 \begin{equation}\label{eq:repn7nf}
 z_{j+1} = -(1+\mu) z_j + P|z_j|^2z_j + Q|z_j|^4z_j + R{\bar z}^5
 \end{equation}
for representation~7 (truncated at quintic order), where the two amplitudes
of~$f_1$ and $f_2$ in (\ref{eq:repn7fns1}--\ref{eq:repn7fns2}) are the real and
imaginary parts of~$z$, and $P$, $Q$ and~$R$ are real constants. For
representation~10 we truncate at cubic order and obtain:
 \begin{align}\label{eq:repn10nf1}
 a_{j+1} &= -(1+\mu) a_j + Pa_j^3 + Q(a_j^2+b_j^2+c_j^2)a_j, \\
 b_{j+1} &= -(1+\mu) b_j + Pb_j^3 + Q(a_j^2+b_j^2+c_j^2)b_j, \\
 c_{j+1} &= -(1+\mu) c_j + Pc_j^3 + Q(a_j^2+b_j^2+c_j^2)c_j, 
 \label{eq:repn10nf3}
 \end{align}
where $P$ and $Q$ are (different) real constants. In these two sets of
equations, $\mu$~represents the bifurcation parameter. The $-1$ Floquet
multipliers at $\mu=0$ arise because these are subharmonic bifurcations. In
representation~7, equivariance with respect to $R_{\tau_T}=-I_2$ is a normal
form symmetry, so even terms up to any order can be removed
from~(\ref{eq:repn7nf}) by coordinate transformations~\cite{refE2}. With
representation~10, the matrix~$-I_3=R_\rho^3$ appears as a spatial symmetry, so
the normal form symmetry is in fact exact, and every solution branch has the
spatio-temporal symmetry $\tau_T\rho^3$, a rotation by~$180^\circ$
followed by time-translation by one period.

The patterns are neutrally stable with respect to translations in the two
horizontal directions, and so also have two Floquet multipliers equal to~$1$
associated with translation modes. We have neglected these as all the patterns
we find are pinned by reflection symmetries that prohibit drifting.

The final stages are to determine the solutions that are created in each of
these bifurcations, their symmetry and stability properties, and to compare 
these with experimental observations. 

\begin{figure}
 \begin{center}
 \centerline{%
 \hbox to0.30\hsize{\hfil(a)\hfil}\hfil
 \hbox to0.30\hsize{\hfil(b)\hfil}\hfil
 \hbox to0.30\hsize{\hfil(c)\hfil}}
 \centerline{% 
 \epsfxsize0.30\hsize\epsffile{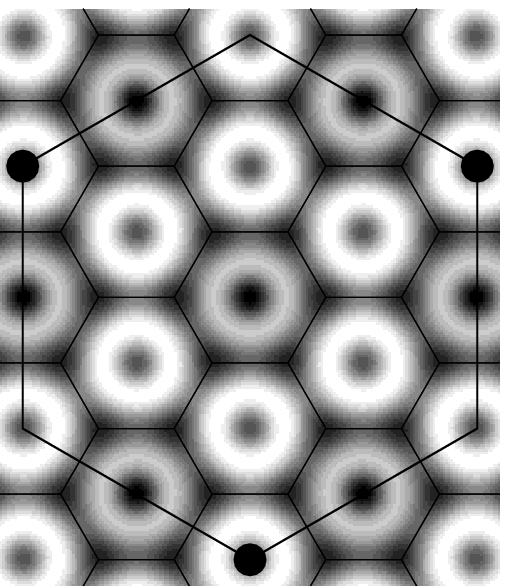}\hfil
 \epsfxsize0.30\hsize\epsffile{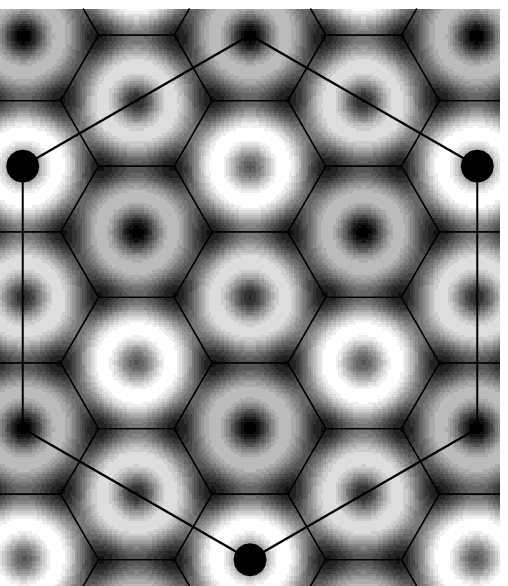}\hfil
 \epsfxsize0.30\hsize\epsffile{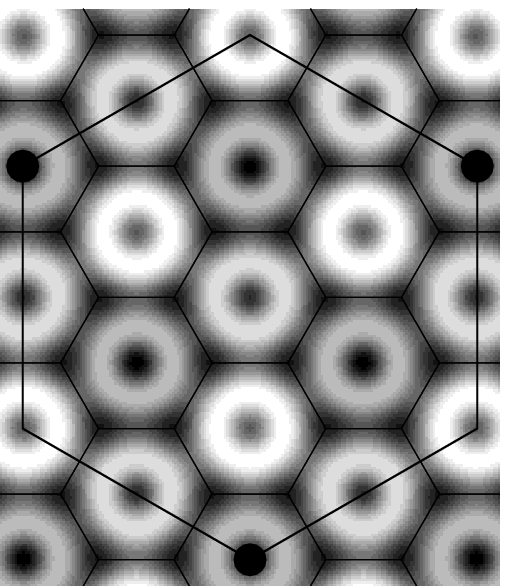}}
 \caption{Reconstructed patterns from the two solutions that arise in
representation~7, using the Fourier functions
(\ref{eq:repn7fns1}--\ref{eq:repn7fns2}) added to a function of the form
of~(\ref{eq:basichexgons}). (a) has the spatial symmetries of pattern~(a) and
no spatio-temporal symmetries (cf.~\ref{fig:idealised}a,d); (b) has the
symmetry properties of pattern~(b) (c is one period~$T$ later;
cf.~figure~\ref{fig:idealised}b,e,g)}
 \label{fig:representation7}
 \end{center}
 \end{figure}

The first normal form (\ref{eq:repn7nf}) generically has two types of
period-two points, found by solving~$f(z)=-z$:
 \begin{equation}
 z_a=          \sqrt{\frac{\mu}{P} - 2\mu^2\frac{Q+R}{P^3}}, \qquad
 z_b=\hbox{i}\,\sqrt{\frac{\mu}{P} - 2\mu^2\frac{Q-R}{P^3}}.
 \end{equation}
The first of these has exactly the symmetry group~$\Sigma_a$ of pattern~(a),
with no spatio-temporal symmetries, while the second has exactly the spatial
symmetry group~$\Sigma_b$ of pattern~(b), as well as spatio-temporal symmetries
generated by~$\rho\tau_T$. Reconstructions of these two are shown in
figure~\ref{fig:representation7}(a) for pattern~(a) and
figure~\ref{fig:representation7}(b,c) for pattern~(b), using the Fourier
functions from above. Linearising the normal form about these two period-two
points readily yields stability information: if $P>0$, then both patterns are
supercritical but only one is stable, while if $P<0$, both are subcritical and
neither is stable.

 \begin{figure}
 \begin{center}
 \centerline{%
 \hbox to0.22\hsize{\hfil(a)\hfil}\hfil
 \hbox to0.22\hsize{\hfil(b)\hfil}\hfil
 \hbox to0.22\hsize{\hfil(c)\hfil}\hfil
 \hbox to0.22\hsize{\hfil(d)\hfil}}
 \centerline{% 
 \epsfxsize0.22\hsize\epsffile{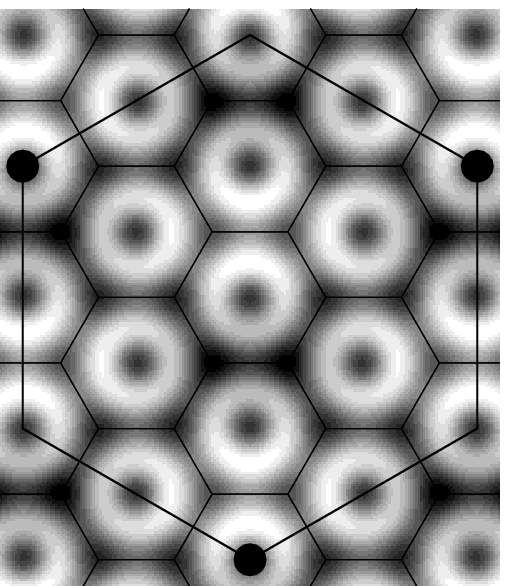}\hfil
 \epsfxsize0.22\hsize\epsffile{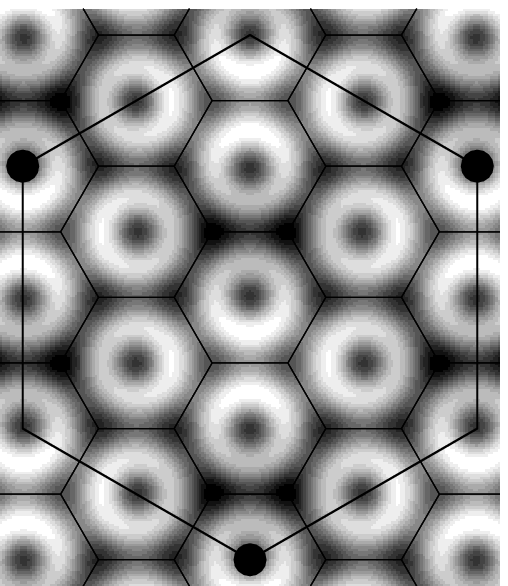}\hfil
 \epsfxsize0.22\hsize\epsffile{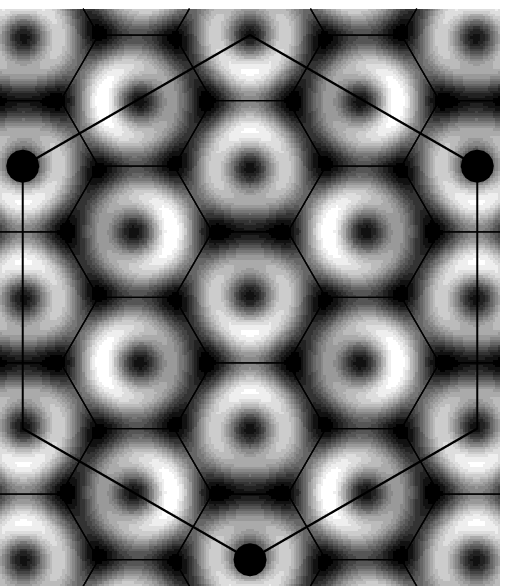}\hfil
 \epsfxsize0.22\hsize\epsffile{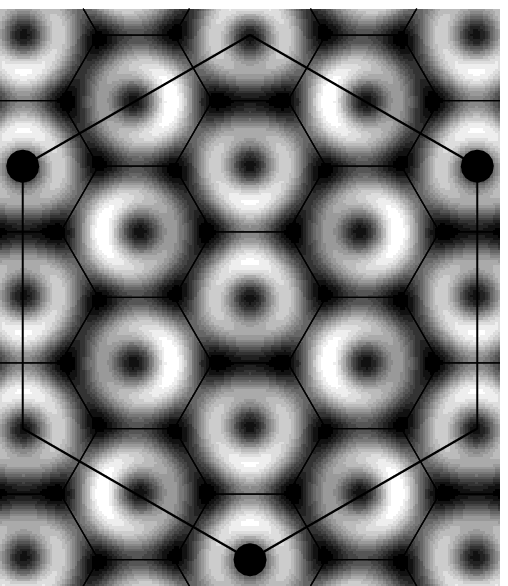}}
 \centerline{%
 \hbox to0.22\hsize{\hfil(e)\hfil}\hfil
 \hbox to0.22\hsize{\hfil(f)\hfil}\hfil
 \hbox to0.22\hsize{\hfil(g)\hfil}\hfil
 \hbox to0.22\hsize{\hfil(h)\hfil}}
 \centerline{% 
 \epsfxsize0.22\hsize\epsffile{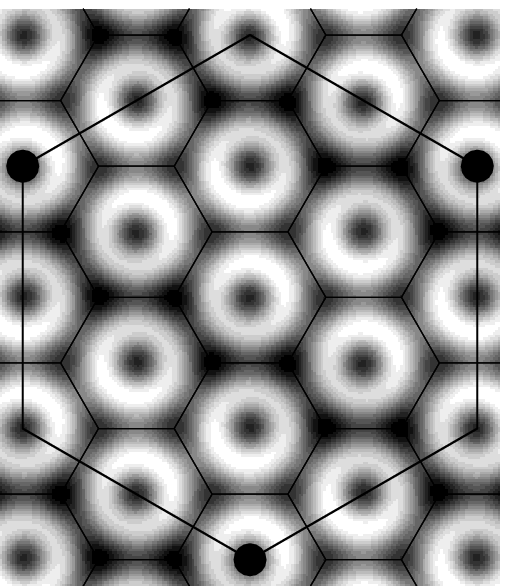}\hfil
 \epsfxsize0.22\hsize\epsffile{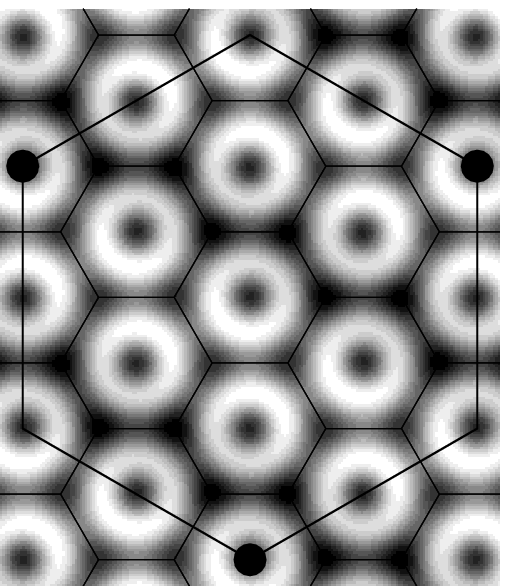}\hfil
 \epsfxsize0.22\hsize\epsffile{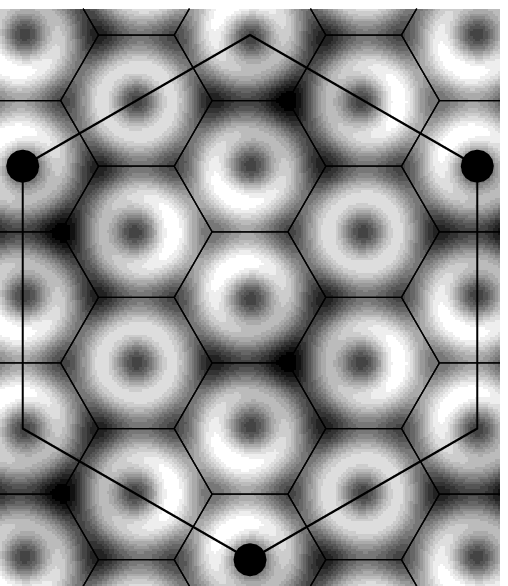}\hfil
 \epsfxsize0.22\hsize\epsffile{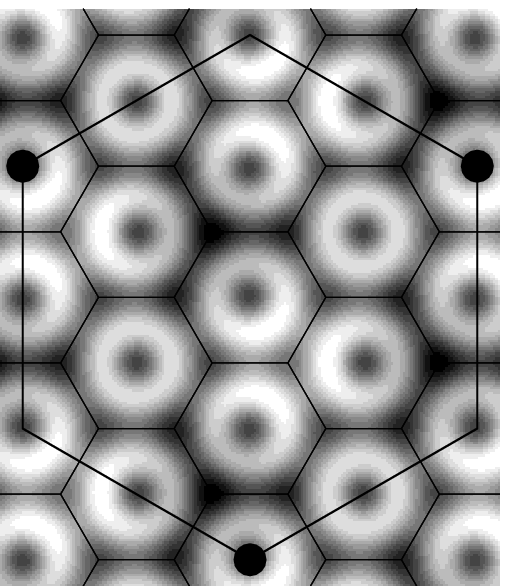}}
 \caption{Reconstructed patterns from irreps~10 and~12: 
 (a,b)~10: $(a,b,c)=(1,1,0)$ (cf.~figure~\ref{fig:idealised}c,f,h);
 (c,d)~12: same amplitudes and same symmetries as (a,b);
 (e,f)~10: $(a,b,c)=(1,0,0)$;
 (g,h)~10: $(a,b,c)=(1,1,1)$.}
 \label{fig:representation1012}
 \end{center}
 \end{figure}

The second normal form (\ref{eq:repn10nf1}--\ref{eq:repn10nf3}) generically has
three types of period-two points~$(a,b,c)$:
 \begin{equation}\label{eq:repn10branches}\quad
 \sqrt{\frac{\mu}{P+Q}}
   \begin{pmatrix}1\\0\\0\end{pmatrix},\quad
 \sqrt{\frac{\mu}{P+2Q}}
   \begin{pmatrix}1\\1\\0\end{pmatrix},\quad
 \sqrt{\frac{\mu}{P+3Q}}
   \begin{pmatrix}1\\1\\1\end{pmatrix}.\quad
 \end{equation}
The middle branch has the spatio-temporal symmetries of pattern~(c), with
12~elements in the spatial part of the symmetry
group~($\Sigma_c=\langle\kappa_x,\kappa_y\tau_2,\tau_1^2\rangle$).
Figure~\ref{fig:representation1012}(a,b) illustrates this pattern
(cf.~figure~\ref{fig:idealised}c,f,h). For comparison, the pattern that would
have been obtained with modes from representation~12 is in
figure~\ref{fig:representation1012}(c,d): the symmetry group is the same, but
the appearance of the pattern does not match the experimental observation. The
first branch has a 24~element spatial symmetry group
$\langle\rho^3\tau_1,\kappa_x\rho\tau_1^5\tau_2,\tau_1^2\rangle$
(figure~\ref{fig:representation1012}e,f), and the third branch has an
18~element group $\langle\kappa_y\tau_2,\kappa_x\rho^5,\tau_1^2\rangle$
(figure~\ref{fig:representation1012}g,h). The three patterns also have the
spatio-temporal symmetry~$\rho^3\tau_T$ (since $R_\rho^3=-I_3$), so $\rho^3$
will appear in the symmetry group of the time-average of each of the
patterns, as discussed in~\cite{refT53}.

The first branch has Floquet multipliers $-1+2\mu$ and $-1-\frac{P}{P+Q}\mu$
(twice); the second branch $-1+2\mu$, $-1-\frac{P}{P+2Q}\mu$ and
$-1+\frac{2P}{P+2Q}\mu$; and the third branch $-1+2\mu$ and
$-1+\frac{2P}{P+3Q}\mu$ (twice). As a result, if $P+Q>0$ and $P+3Q>0$, then all
branches bifurcate supercritically, and either the first branch will be stable
(when~$P<0$) or the last will be stable (when~$P>0$). If any branch bifurcates
subcritically, none are stable. The middle branch, which is the one
corresponding to the experimentally observed pattern~(c), is always unstable at
onset. 
%This is discussed in more detail below.

\section{Discussion}\label{sec:discussion}
Using the symmetry-based approach of Tse \etal~\cite{refT53}, we have
analysed three experimentally observed spatial period-multiplying transitions
from an initial hexagonal pattern. The three patterns illustrate three
situations that can arise in this kind of analysis. Pattern~(a) was
straight-forward, in that a single representation of~$\Gamma$ had a
one-dimensional space fixed by the spatial symmetry group of the pattern. The
existence of a solution branch of the form of pattern~(b) could also be
inferred using the Equivariant Branching Lemma, though in this case it was
necessary to include the temporal symmetry associated with period-doubling
bifurcation. Specifically, the spatial symmetries selected a two-dimensional
fixed point space which was further reduced to a one-dimensional fixed point
space when spatio-temporal symmetries were taken into account. Experimentally,
these two patterns were found for the same fluid parameters and same
$2\omega:3\omega$ forcing function but for different frequencies~$\omega$:
$\omega=25\,\hbox{Hz}$ for~(a) and $\omega=35\,\hbox{Hz}$ for~(b). This
suggests that the transition between these patterns, which arise for
instabilities associated with the same representation, might be observed by
tuning the frequency~$\omega$.

Pattern~(c), on the other hand, had a spatial symmetry group that fixed
one-dimensional subspaces in two different representations, and we appealed to
the measured power spectrum of the pattern to choose between the two
possibilities. In this situation, symmetry considerations alone were not
enough. Similar situations arise in other bifurcation problems, for example,
knowing that a stable axisymmetric pattern is found in a spherically symmetric
bifurcation problem does not provide enough information to determine which is
the relevant representation. 

The experimentally observed transition between hexagons and pattern~(c) occurs
by means of a propagating front that separates domains of hexagons and the
secondary pattern. The front is initiated at the lateral boundaries of the
system and emanates radially inward. There is little if any hysteresis, and the
reverse transition also occurs via the same scenario. The occurrence of a front
in this transition suggests bistability of the hexagonal pattern and
pattern~(c).  This is certainly consistent with the theoretical prediction that
pattern~(c) is unstable at small amplitude, that is, at onset. However, we have
not explored the possible stabilization mechanisms for pattern~(c).

It is worth emphasizing that an understanding of group representation theory is
useful in classifying and analysing secondary instabilities of patterns,
not only in the Faraday wave experiment as described here, but also in
convection and other pattern formation problems (see~\cite{refR59}). It
is also worth mentioning that the examples studied here indicate that
spatio-temporal symmetries readily arise in secondary subharmonic
instabilities, and that careful experimental characterization of these, either
by still images taken one forcing period apart or by time-averaging over two
forcing periods, can be helpful. Subsequent instabilities of patterns that have
spatio-temporal symmetries can be analysed using methods described
in~\cite{refR40,refL51}.

The approach outlined in~\cite{refT53} and here is useful for taking an
experimental observation of a secondary transition and casting it into its
equivariant bifurcation theory context, but it does not predict which
transitions should be expected in an experiment. However, in these
two-frequency Faraday wave experiments, three-wave interactions of the type
described in~\cite{refS105} may select a third wavevector that could appear in
the secondary transition. Each of the representations in the problem under
consideration is associated with a set of wavevectors, providing a possible
mechanism for selecting between possibilities.

\noindent{\bf Acknowledgements.} This paper builds on earlier published results
obtained with Dawn Tse, Rebecca Hoyle and Hagai Arbell. AMR is grateful for
support from the EPSRC. The research of MS is supported in part by NSF grant
DMS-9972059 and NASA grant NAG3-2364. JF is grateful for support from the
Israel Academy of Science (grant 203/99).

\end{document}